\documentclass[reprint, 10pt, a4paper, aps, prx, floatfix, longbibliography]{revtex4-1}
\usepackage[utf8]{inputenc}
\usepackage{geometry} 
\usepackage{graphicx} 
\usepackage{amsmath, amsbsy}
\usepackage{amssymb}
\usepackage{float} 
\usepackage{wrapfig} 
\usepackage[hidelinks]{hyperref}

\newcommand{\bx}{\mathbf{x}}

\newcommand{\D}{\mathrm{d}}

\begin{document}
\title{Recovering Gardner Restacking with Purely Diffusive Operations}
\date{\today}
\author{E. J. Kolmes}
\email{ekolmes@princeton.edu}
\author{N. J. Fisch}
\affiliation{Department of Astrophysical Sciences, Princeton University, Princeton, New Jersey, USA}

\begin{abstract}
The maximum particle kinetic energy that can be extracted from an initial six-dimensional phase space distribution motivates the concept of free or available energy. 
The free energy depends on the allowed operations that can be performed. 
A key  concept underlying the theoretical treatment of plasmas is the Gardner free energy, where the exchange of the contents of equal phase volumes is allowed. 
A second  free energy concept is the diffusive free energy, in which the contents of volumes are instead averaged. 
For any finite discretization of phase space, the diffusive free energy is known to be less than the Gardner free energy. 
However, despite the apparent fundamental differences between these free energies, it is demonstrated here that the Gardner free energy may be recovered from the continuous limit of the diffusive free energy, leading to the surprise that macroscopic phase-space conservation can be achieved by ostensibly entropy-producing microscopic operations. 
\end{abstract}

\maketitle

\section{Introduction}


One of the key problems in the field of plasma physics is the instability of the plasma, which then begets the question, if the plasma is unstable, then how much energy can be released.
A very important subset of these instabilities is when the energy released is the kinetic energy of charged particles.
This release of particle kinetic energy may be framed as the reorganization of the particle phase space in which high energy particles are reorganized to occupy lower energy states.  
The release is stimulated by wave-particle interactions, whereby when the particle energy is released, the wave grows in amplitude.
In practical devices, often these instabilities are deleterious, but sometimes they are advantageous.


The release of this energy depends on the nature of the allowed wave-particle interactions.
The {\it free energy} can be thought of as the maximum energy available for release subject to constraints.
The most mild constraint, first presented by Gardner \cite{Gardner1963}, applies for any system in which densities of volumes in six-dimensional phase space are conserved.
Let phase space be divided into small, discrete regions of constant phase-space volume, so that the conservation constraint requires that phase space can only be reconfigured by pairwise exchange of the densities in these regions. 
Once the phase space is rearranged so that the highest-density volumes are assigned to the lowest-energy regions in phase space, it is not possible to extract further energy while still preserving the phase space densities. 
The process of rearranging the phase space volumes so that density is anticorrelated with energy is often called {\it Gardner restacking}. 
By construction, the {\it Gardner free energy} obtained by Gardner restacking is the maximum energy that can be released under the mildest of constraints.

However, when the plasma distribution function is viewed with any finite granularity, many processes can appear to diffuse particles between volumes of phase space rather than exchanging the contents of individual volumes \cite{Kennel1966, Fisch1992}. 
As a result, it is often useful to consider an alternative to Gardner's problem, where the maximum accessible energy is determined by what can be extracted by diffusion between phase space volumes (including elements which are not adjacent) rather than Gardner restacking \cite{Fisch1993, Hay2015, Hay2017, Kolmes2020ConstrainedDiffusion}. 
This is a qualitatively different process from the pairwise exchange of phase space densities that underlies Gardner restacking; for one thing, every diffusive step creates entropy, whereas restacking is reversible. 

Both the Gardner restacking problem and the diffusive exchange problem have continuous and discrete variants. That is, phase space can be considered continuous or can be divided into discrete regions (either because the system is intrinsically discrete or to represent averaging over those regions). 
The continuous diffusive problem, as posed by Fisch and Rax in 1993 \cite{Fisch1993}, is to minimize 
\begin{gather}
W_\text{final} = \lim_{t \rightarrow \infty} \int \varepsilon(v) f(v,t) \, \D v
\end{gather}
by evolving the distribution $f(v,t)$ according to 
\begin{gather}
\frac{\partial f}{\partial t} = \int K(v,v',t) \big[ f(v',t) - f(v,t) \big] \D v' .
\end{gather}
Here $\varepsilon(v)$ is the energy per particle and the kernel $K$ must satisfy $K(v,v',t) = K(v',v,t)$ and $K(v,v',t) \geq 0$. Fisch and Rax showed that $f$ satisfies an $H$ theorem, and that the system will reach a steady state, but they  left the matter of the releasable free energy as an open problem, noting that it is ``quite formidable" given the necessity to search over all possible kernels $K(v,v',t)$.
Indeed, in the years since, substantial progress has been made on the discrete diffusive exchange problem \cite{Hay2015, Hay2017, Kolmes2020ConstrainedDiffusion}, as well as on continuous Gardner restacking \cite{Dodin2005, Helander2017ii, Helander2020}, but the minimum energy state under continuous diffusive operations remains unsolved. 

This paper will show that, in fact, in the continuous limit, the free energy available under the diffusive constraint is equivalent to the Gardner free energy under the restacking constraint. 
This is a counterintuitive result: one problem involves purely reversible operations and the other involves irreversible operations. 
%
%
In proving this equivalence, this paper now also solves for the minimum energy state under diffusion in the continuous limit, a problem previously considered intractable. 
In addition, in proving this equivalence, this paper provides a prescription for constructing a kernel $K(v,v',t)$ that approaches the minimum energy arbitrarily closely.
%



Recognizing this equivalence provides new intuitions regarding the broader nature of phase space granularity and irreversible operations, which may be of interest to communities beyond the field of plasma physics.
Although Gardner restacking had been proposed to quantify the free energy in plasma, 
the underlying concepts can be applied to a variety of settings outside of plasma physics \cite{Berk1970, Bartholomew1971, Lemou2012, Chavanis2012, Baldovin2016}. 
Interestingly, the same concept has been treated within the  mathematical literature, where the equivalent of a Gardner restacked distribution is called the ``symmetric decreasing rearrangement" of a function \cite{Riesz1930, HardyLittlewoodPolya, Brascamp1974, Almgren1989, Baernstein}.

Also enjoying  broad interest is the free energy under the diffusion constraint.  
The  free energy  in plasma through diffusion by waves is of practical interest in channeling energy from the byproducts of the nuclear fusion reaction in controlled magnetic confinement fusion.   
A variety plasma waves at different frequencies have been proposed to accomplish this diffusion
\cite{Valeo1994, Fisch1995ii, Ochs2015a, Ochs2015b, Chen2016,Cianfrani2018,Cianfrani2019,Castaldo2019,Romanelli2020} 
as well as combinations of these plasma waves \cite{Fisch1995, Herrmann1997}. 
But the discrete diffusive problem also appears in a number of contexts outside of plasma physics, including physical chemistry \cite{Zylka1985},  income inequality \cite{Dalton1920, Atkinson1970, Aboudi2010}, and altruism \cite{Thon2004}. 
The general problem of determining the states accessible via an allowable set of operations has wide applicability -- from chemistry \cite{Horn1964} to laser physics \cite{Levy2014} to quantum information theory and thermodynamics \cite{Lostaglio2015, Brandao2015, Korzekwa2019}. 

The paper is organized as follows: 
Section~\ref{sec:discreteRules}  describes the discrete restacking and diffusion models and some of their properties. 
Section~\ref{sec:sequenceProof} shows how restacking and diffusion converge to the same behavior in the continuous limit. 
Section~\ref{sec:entropy} shows how the entropy production associated with diffusive exchanges can be suppressed in the continuous limit. 
Section~\ref{sec:scales} discusses issues related to characteristic scales in phase space.
Section~\ref{sec:bumpOnTail} applies these results to the classic bump-on-tail distribution.
Section~\ref{sec:discussion} presents a summary and broader discussion of the results.

\section{Discrete Restacking and Diffusion} \label{sec:discreteRules}

Consider a distribution $f$ that is a function of a phase space coordinate (or vector of coordinates) $\bx$. Suppose a particle at $\bx$ has energy $\varepsilon(\bx)$. Moreover, suppose $f$ is piecewise continuous and $\varepsilon$ is Riemann integrable. If the domain of $\bx$ is divided into some set of equal-volume regions $\{S_i\}$, define 
\begin{gather}
f_i \doteq \int_{S_i} f(\bx) \, \D \bx 
\end{gather}
and 
\begin{gather}
\varepsilon_i \doteq \int_{S_i} \varepsilon(\bx) \, \D \bx . 
\end{gather}
Then the discrete Gardner restacking problem consists of exchanging the $\{f_i\}$ in order to minimize $\sum_i \varepsilon_i f_i$ and the discrete diffusive problem consists of averaging pairs $f_i$ and $f_j$ to minimize the same expression. The continuous restacking and diffusion problems can be viewed as the infinitely fine-grained limits of the corresponding discrete problems. 

There are three things to note about the diffusive exchange operation. First, $f_i$ and $f_j$ need not correspond to adjacent regions in phase space in order to be averaged; the operation can be macroscopically non-local. Microscopically local dynamics can give rise to exchanges of material between non-contiguous regions of phase space \cite{Fisch1993}. 

Second, the diffusive free energy is defined as the \textit{maximal} energy that can be extracted from an initial distribution. If a ground state is any state from which no further energy can be extracted, different sequences of diffusive exchange operations on the same initial state can lead to different ground states with different energies. Finding the diffusively accessible free energy is an optimization problem on the space of sequences of diffusive exchanges. This is a large part of why the diffusive problem tends to be analytically more difficult than the restacking problem. 

Finally, the energy that can be released through diffusive exchanges never exceeds the Gardner free energy, and the two are only exactly equal when both vanish (that is, when the distribution begins in a ground state). 
To see this, note that both processes produce final distribution functions in which the most populated volumes of phase space are assigned to the lowest-energy regions of phase space.
The discrepancies in the populations create the opportunities to release energy.
However, each diffusive exchange reduces the difference between the high-population and low-population volumes, leaving less opportunity for reducing the energy in the final state. 

\section{Recovering Gardner Restacking with Diffusive Operations} \label{sec:sequenceProof}

The basic operation of Gardner restacking is to exchange the populations of two equal-volume regions. For a sufficiently fine discretization of phase space, $f(\bx)$ can be considered constant over each region. 
Suppose phase space is then further subdivided into even smaller regions. 
Then, as will be shown here,  it is possible to use diffusive exchange operations on this finer grid to approach the results of the original Gardner restacking operation on the coarser grid. 
In the limit of an arbitrarily fine discretization, the diffusive operations can approach this limit arbitrarily closely. 

To show this, consider two regions of phase space, $A$ and $B$, each with volume $V$. Both Gardner restacking and diffusive exchange operations act only on the difference between two populations, so for the sake of simplicity (and without loss of generality) assume that region $A$ initially has population density $f_A = 0$ and region $B$ initially has population density $f_B = f_0$. Then a Gardner restacking operation between regions $A$ and $B$ would exchange the populations, so that $f_A = f_0$ and $f_B = 0$. 

Now consider a subdivision of $A$ and $B$ each into $N$ regions with volume $V / N$. Let $a_i f_0$ and $b_i f_0$ denote the population densities of the $i$th sub-regions within $A$ and $B$, respectively. If $f(\bx)$ was originally constant over the regions $A$ and $B$, then before any diffusive exchanges, $a_i = 0$ and $b_i = 1$ $\forall i$. 

To move the contents of $B$ to $A$, perform the following sequence of diffusive exchanges: first $a_0$ with $b_0$, then $a_0$ with $b_1$, and so on until $a_0$ exchanges with $b_{N-1}$. Next, perform the same exchanges but with $a_1$ instead of $a_0$, and so on for each $a_i$, until the final exchange is $a_{N-1}$ with $b_{N-1}$. In total, there will be $N^2$ diffusive exchange operations. 
For the sake of concreteness, it may be useful to visualize this process for small $N$. When $N=1$, Gardner restacking takes 
\begin{gather*}
\begin{array}{|c|}
\hline 0 \\ \hline
\end{array} \;
\begin{array}{|c|}
\hline 1 \\ \hline
\end{array}
\rightarrow
\begin{array}{|c|}
\hline 1 \\ \hline
\end{array} \;
\begin{array}{|c|}
\hline 0 \\ \hline
\end{array} \, ,
\end{gather*}
whereas diffusive exchange takes 
\begin{gather*}
\begin{array}{|c|}
\hline 0 \\ \hline
\end{array} \; \;
\begin{array}{|c|}
\hline 1 \\ \hline
\end{array}
\rightarrow
\begin{array}{|c|}
\hline 1/2 \\ \hline
\end{array} \; \;
\begin{array}{|c|}
\hline 1/2 \\ \hline
\end{array} \, .
\end{gather*}
Half of the content in $B$ is transferred to $A$. Now, when $N=2$, restacking operations can again fully transfer the populations:
\begin{gather*}
\begin{array}{|c|c|}
\hline 0 & 0\\ \hline
\end{array} \; \;
\begin{array}{|c|c|}
\hline 1 & 1\\ \hline
\end{array}
\rightarrow
\begin{array}{|c|c|}
\hline 0 & 1\\ \hline
\end{array} \; \;
\begin{array}{|c|c|}
\hline 1 & 0\\ \hline
\end{array}
\rightarrow
\begin{array}{|c|c|}
\hline 1 & 1\\ \hline
\end{array} \; \;
\begin{array}{|c|c|}
\hline 0 & 0\\ \hline
\end{array} \, ,
\end{gather*}
whereas the sequence of diffusive moves described above does as follows: 
\begin{align*}
\begin{array}{|c|c|}
\hline 0 & 0\\ \hline
\end{array} \; \;
\begin{array}{|c|c|}
\hline 1 & 1\\ \hline
\end{array}
&\rightarrow
\begin{array}{|c|c|}
\hline 1/2 & 0 \\ \hline
\end{array} \; \;
\begin{array}{|c|c|}
\hline 1/2 & 1 \\ \hline
\end{array} \\
&\hspace{0 pt}
\rightarrow 
\begin{array}{|c|c|}
\hline 3/4 & 0 \\ \hline
\end{array} \; \;
\begin{array}{|c|c|}
\hline 1/2 & 3/4 \\ \hline 
\end{array} \\
&\hspace{0 pt}
\rightarrow 
\begin{array}{|c|c|}
\hline 3/4 & 1/4 \\ \hline
\end{array} \; \;
\begin{array}{|c|c|}
\hline 1/4 & 3/4 \\ \hline 
\end{array} \\
&\hspace{0 pt}
\rightarrow 
\begin{array}{|c|c|}
\hline 3/4 & 1/2 \\ \hline
\end{array} \; \;
\begin{array}{|c|c|}
\hline 1/4 & 1/2 \\ \hline 
\end{array} \, .
\end{align*}
The prescribed sequence of moves transfers $5/8$ of the total population from $B$ to $A$. 
Note that $5/8$ is already greater than $1/2$, meaning that it has already been shown in this simple example that a sequence of diffusive operations can achieve non-diffusive behavior when viewed on a coarser scale.

More generally, and more formally, let $a_i^{(s)}$ and $b_i^{(s)}$ denote the values of $a_i$ and $b_i$ after the first $s \cdot N$ exchanges (in other words, immediately after the exchange between $a_s$ and $b_{N-1}$). The final value of $a_i$ will be fixed by the last exchange involving $a_i$, so after all exchanges, $a_i = b_{N-1}^{(i+1)}$. The objective is to prove that the entire content of $B$ can be transferred to $A$ in the limit of large $N$, a statement which can be rewritten as 
\begin{gather}
\lim_{N \rightarrow \infty} \frac{1}{N} \sum_{i=0}^{N-1} b_{i}^{(N)} \stackrel{?}{=} 0. \label{eqn:desiredSum}
\end{gather}

The value of any $b_i^{(s+1)}$ can be written recursively in terms of the values of $a_{s}^{(0)}$ and $b_j^{(s)}$. In particular, 
\begin{gather}
b_i^{(s+1)} = \sum_{j=0}^{N-1} M_{ij} b_j^{(s)} + 2^{-i-1} a_i^{(0)}, \label{eqn:recursion}
\end{gather}
where $M_{ij}$ is the lower triangular Toeplitz matrix given by 
\begin{gather}
M_{ij} = \begin{cases}
2^{-i+j-1} & i \geq j \\
0 & i < j. 
\end{cases}
\end{gather}
For the initial conditions in this scenario, $a_i^{(0)} = 0$, so the corresponding term in Eq.~(\ref{eqn:recursion}) can be ignored. Then if $(M^s)_{ij}$ denotes the $ij$ element of the matrix $M$ to the $s$th power, Eq.~(\ref{eqn:recursion}) becomes 
\begin{gather}
b_i^{(s)} = \sum_{j=0}^i (M^s)_{ij} b_j^{(0)}. 
\end{gather}
For $i < j$, $(M^s)_{ij} = 0$. It can be shown by induction on $s$ that the nonzero elements are 
\begin{align}
&(M^s)_{ij} = 2^{-i+j-s} \frac{\Gamma(i-j+s)}{\Gamma(i-j+1) \Gamma(s)}
&(i \geq j),
\end{align}
where $\Gamma$ is the usual gamma function. Then 
\begin{align}
b_i^{(s)} = \sum_{j=0}^i 2^{-i+j-s} \frac{\Gamma(i-j+s)}{\Gamma(i-j+1) \Gamma(s)} \, b_j^{(0)}. \label{eqn:initialConditionMapping}
\end{align}
Using the initial condition that $b_j^{(0)} = 1$, it follows (after some manipulation) that 
\begin{align}
\sum_{i=0}^{N-1} b_i^{(N)} &= \sum_{i=0}^{N-1} \sum_{k=0}^{i} 2^{-N-k} \frac{\Gamma(N+k)}{\Gamma(k+1) \Gamma(N)} \\
&= \sum_{k=0}^{N-1} \frac{N-k}{2^{N+k}} \frac{\Gamma(N+k)}{\Gamma(k+1) \Gamma(N)} \, .
\end{align}
It is possible to show, using induction on $M$, that 
\begin{gather}
\sum_{k=0}^M \frac{N-k}{2^k} \frac{\Gamma(N+k)}{\Gamma(k+1)} = \frac{1}{2^M} \frac{\Gamma(N+M+1)}{\Gamma(M+1)} \, .
\end{gather}
Therefore, 
\begin{align}
\frac{1}{N} \sum_{i=0}^{N-1} b_{i}^{(N)} &= \frac{1}{2^{2N-1}} \frac{\Gamma(2N)}{\Gamma(N) \Gamma(N+1)} \, . \label{eqn:bSum}
\end{align}
Applying Stirling's approximation, the large-$N$ limit is 
\begin{align}
\lim_{N \rightarrow \infty} \frac{1}{N} \sum_{i=0}^{N-1} b_{i}^{(N)} &= \sqrt{ \frac{1}{\pi N} } + \mathcal{O} \big( N^{-3/2} \, \big) . \label{eqn:NScaling}
\end{align}
This is sufficient to prove Eq.~(\ref{eqn:desiredSum}): in the large-$N$ limit, an arbitrarily large fraction of the population will be transferred from $B$ to $A$. Equation~(\ref{eqn:NScaling}) is a major result of this paper. 

\begin{figure}
\centering
\includegraphics[width=\linewidth]{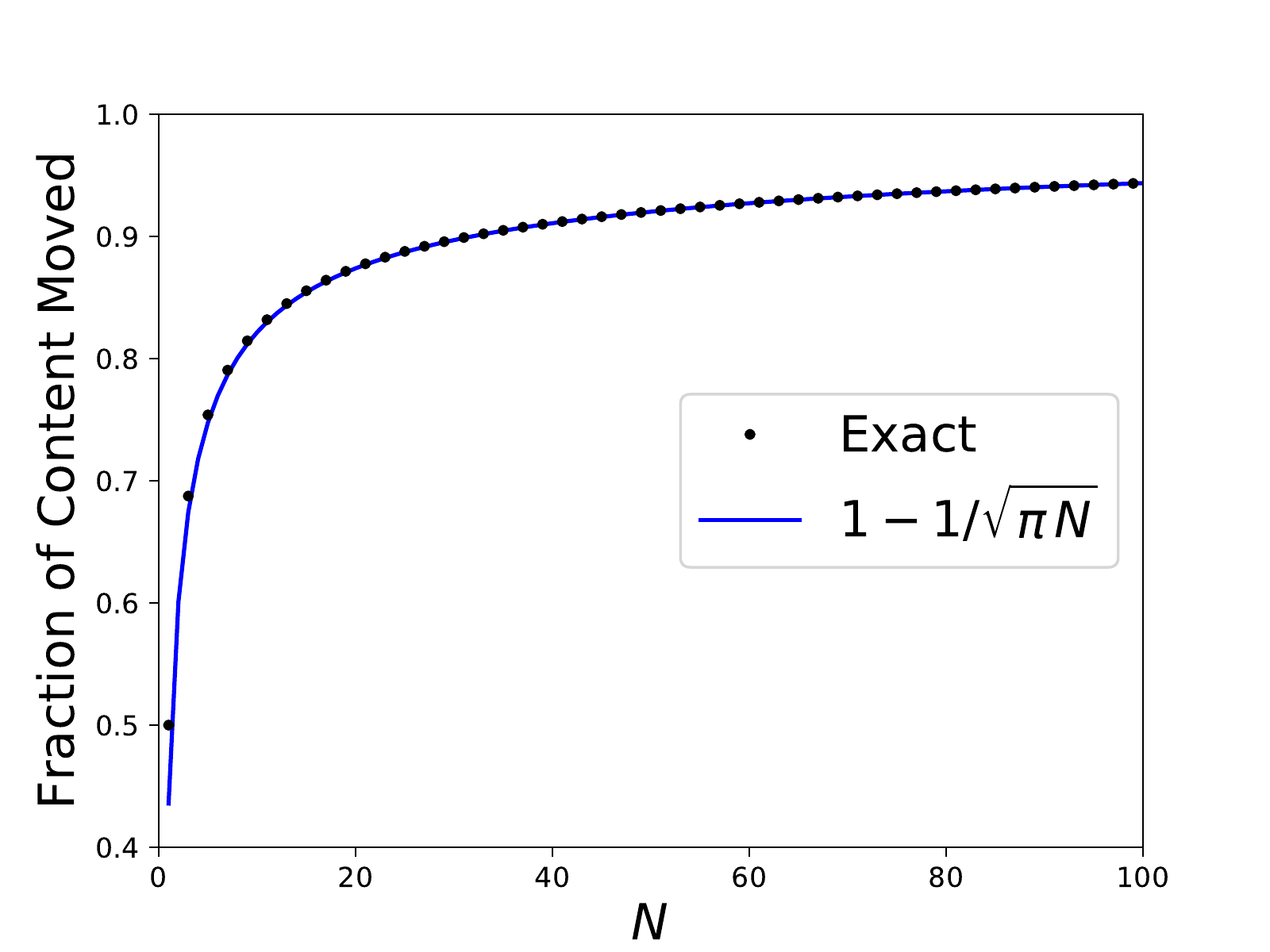}
\caption{Fractional content transfer between the two regions vs.~$N$.  Exact transfer (following the sequence of diffusive exchanges described in Section~\ref{sec:sequenceProof}) as a function of $N$ is well-approximated by $1-1/\sqrt{\pi N}$.} \label{fig:transferScaling}
\end{figure}

Note that this proof has not demonstrated that this particular sequence of diffusive exchanges is optimal. 
Therefore, Eq.~(\ref{eqn:NScaling}) provides a lower bound for how quickly the diffusive free energy can approach the Gardner free energy as the scale becomes  finer.
In principle there could be a sequence that could do so in fewer steps. 
Of course, the main point here is the fact that the large-$N$ limit does converge to complete population transfer, which implies that the optimal sequence must also similarly converge. 
This is sufficient to show the counterintuitive and rather remarkable result that, for a piecewise continuous initial distribution, the free energy under the continuous diffusive exchange constraint is the same as that under Gardner restacking in the continuous limit. 
%

Also, note that in providing a prescription for releasing the Gardner free energy, an upper bound is established on the number of steps required to realize this energy release to any required accuracy.  To see this, divide the phase space first into $M$ phase space volumes.  
The Gardner restacking requires sorting these $M$ volumes, which can be accomplished in $\mathcal{O}(M \ln M)$ pairwise exchanges. 
Now upon a further subdivision of each of the $M$  volumes to $N$ sub-volumes, and executing instead $N^2$ diffusive steps, each of those exchanges can be accomplished to accuracy $N^{-1/2}$.
Thus, the Gardner free energy to accuracy $N^{-1/2}$ may be realized in no greater than $\mathcal{O}(N^2 M \ln M)$ pairwise diffusive exchanges.

\section{Entropy and Reversibility} \label{sec:entropy}

The previous section demonstrates how a series of irreversible operations can mimic the behavior of a reversible operation. 
It is clear that the entropy production must somehow be suppressed as $N$ becomes large. 
To show how this happens, we track when entropy is created and destroyed as the procedure described in Section~\ref{sec:sequenceProof} is followed. 

Consider an entropy defined by 
\begin{gather}
S = - \sum_i p_i \log p_i ,
\end{gather}
where $p_i$ is the probability that a particle will occupy state $i$. For  simplicity, pick a normalization such that initially, $p_A = 0$ and $p_B = 1$ (where $A$ and $B$ are the two regions described in Section~\ref{sec:sequenceProof}). 

Subdividing states changes the entropy $S$. If each state is divided into $N$ states with probability $p_i / N$, entropy increases by 
\begin{gather}
\Delta S = \log N. \label{eqn:fineGridEntropy}
\end{gather}
Each of the subsequent diffusive averaging operations also creates entropy. If states with probabilities $p_i$ and $p_j$ are averaged, the increase in entropy is 
\begin{align}
\Delta S &= - (p_i + p_j) \log \bigg( \frac{p_i+p_j}{2} \bigg) \nonumber \\
&\hspace{70 pt}+ p_i \log p_i + p_j \log p_j  \, . \label{eqn:smallDiffusionEntropyNoDelta}
\end{align}
This is always non-negative. If $p_j / p_i = 1 + \delta$, Eq.~(\ref{eqn:smallDiffusionEntropyNoDelta}) can be written as 
\begin{align}
\Delta S = p_i \bigg[ \frac{\delta^2}{4} + \mathcal{O}(\delta^3) \bigg]. \label{eqn:smallDiffusionEntropy}
\end{align}
After all of the diffusive steps, transforming back from the finer discretization to the coarser one will then destroy entropy. In the limit where each of the coarse-grid states will be constructed out of $N$ identical fine-grid ones, this will exactly cancel the entropy production given in Eq.~(\ref{eqn:fineGridEntropy}). If the fine-grid states are not all equal, it will destroy somewhat less entropy than was created when the coarse-grid states were subdivided. 

\begin{figure}
\centering
\includegraphics[width=\linewidth]{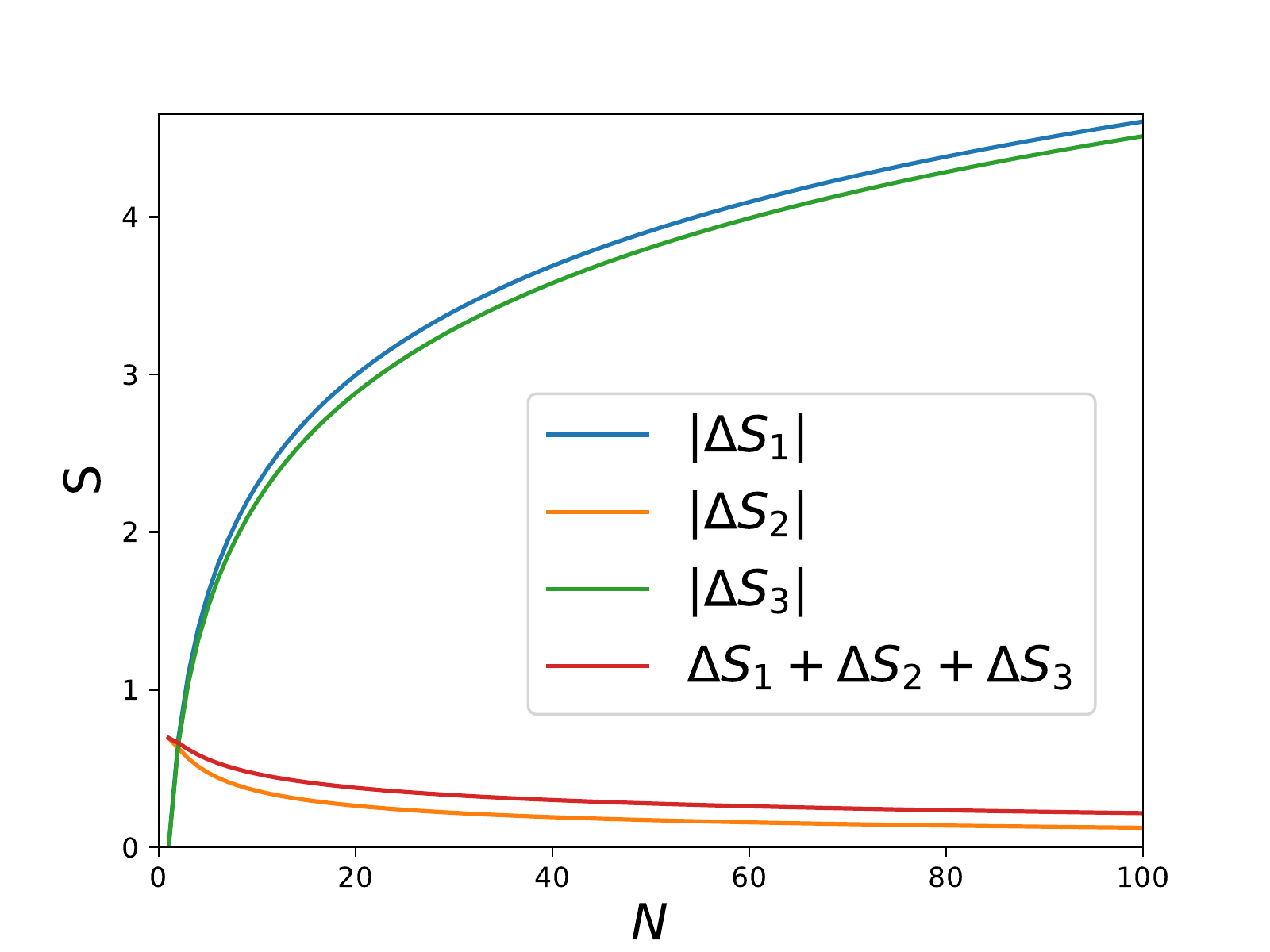}
\caption{Scaling of the different entropy terms with $N$.} \label{fig:entropyScaling}
\end{figure}

For the procedure described in Section~\ref{sec:sequenceProof}, denote the entropy production due to subdividing the states by $\Delta S_1$; denote the total entropy produced by diffusive averaging operations by $\Delta S_2$; and denote the change in entropy when the resulting states are recombined back to the coarser discretization by $\Delta S_3$. 
As $N \rightarrow \infty$, the procedure is able to replicate a reversible exchange, so it must be true that $\Delta S_1 + \Delta S_2 + \Delta S_3 \rightarrow 0$. 
In fact, $\Delta S_1$ and $\Delta S_3$ will cancel one another, and $\Delta S_2$ will vanish on its own. 
The reason for this essentially follows from the quadratic scaling given in Eq.~(\ref{eqn:smallDiffusionEntropy}); increasing $N$ results in a larger number of diffusive operations between states with more similar populations. 
The scaling of these three terms is shown for $N \leq 100$ in Figure~\ref{fig:entropyScaling}. 

It is interesting to note that the composition of operations, that is, the transformations between grids and the averaging steps, is quite similar to the splitting methods used in structure-preserving geometric algorithms \cite{Qin2013, He2015, Morrison2017, Glasser2020}. 
In structure-preserving algorithms, higher-order algorithms of a system can be composed of solutions of subsystems, each of which may not be an algorithm for the full system. 
Similarly, the composition of the microscopic operations proposed here respects properties over a coarser structure that are not respected by the microscopic operations individually.

\section{Characteristic Scales} \label{sec:scales}

As the previous sections  demonstrated, entropy production is strongly dependent on the granularity of the distribution function. 
Subdividing a discretization of the distribution function into successively finer pieces makes the optimal sequence of diffusive operations produce vanishingly little entropy. 
This leads to a question: what does it mean for a discretization to be fine enough? 
In other words, is there some characteristic scale with which to compare the scale of a grid? 

Of course, if the objective is to use the finer grid to reproduce Gardner restacking on the coarser grid, there are conditions that must be met in order for the coarse discretization to be a reasonable approximation of the continuous distribution function in the first place. In particular, if $L$ is a characteristic length of the coarser discretization, $f$ should not vary too much over that length scale. If $f$ is smooth, that condition could be written as $L |\nabla f| \ll f$. 

If $f$ is constant over each discrete region, then it follows from Section~\ref{sec:sequenceProof} that the only further requirement is that $N \gg 1$; the net efficiency of the transfer between the regions scales like $1/\sqrt{\pi N}$. However, one might imagine that the scaling could be different if $\nabla f \neq 0$. 

Consider a generalization of the initial conditions from Section~\ref{sec:sequenceProof} in which $f$ has some gradient over the region $B$. Suppose it is still flat over $A$, so that after $A$ and $B$ are subdivided, 
\begin{gather}
a_i^{(0)} = 0 \\
b_i^{(0)} = 1 + \frac{L f_0'}{N f_0} \bigg( i - \frac{N-1}{2} \bigg),
\end{gather}
where $f_0'$ is constant and $L$ is the characteristic size of the region $B$. This choice of $b_i^{(0)}$ retains the property that $\sum_i b_i^{(0)} = N$. 

The calculation proceeds identically with the new initial conditions through Eq.~(\ref{eqn:initialConditionMapping}). $b_i^{(s+1)}$ is a linear function of each of the $b_j^{(0)}$, so the corrections due to the gradient can be calculated independently. In particular, the correction to Eq.~(\ref{eqn:bSum}) is 
\begin{align}
&\frac{1}{N} \sum_{i=0}^{N-1} \sum_{j=0}^i 2^{-i+j-N} \frac{\Gamma(i-j+N)}{\Gamma(i-j+1) \Gamma(N)} \, (b_j^{(0)}-1) \nonumber \\
&\hspace{20 pt}= \bigg[ - \frac{1}{2} + \frac{(N+1) \Gamma(2N)}{2^{2N} \Gamma(N) \Gamma(N+1)} \bigg] \frac{L f_0'}{N f_0} .
\end{align}
Including the correction, 
\begin{align}
&\lim_{N \rightarrow \infty} \frac{1}{N} \sum_{i=0}^{N-1} b_i^{(N)} \nonumber \\
&\hspace{40 pt}= \sqrt{\frac{1}{\pi N}} \bigg[ 1 + \frac{1}{2} \frac{L f_0'}{f_0} \bigg] + \mathcal{O}(N^{-1}) .
\end{align}
So long as $L$ was chosen to be small enough for the coarser discrete system to be a reasonable approximation of the continuous system -- more precisely, so long as $L \ll f_0 / f_0'$ -- the scaling of the convergence for this sequence of diffusive exchanges is the same to within a small correction.


\begin{figure}
	\centering
	\includegraphics[width=\linewidth]{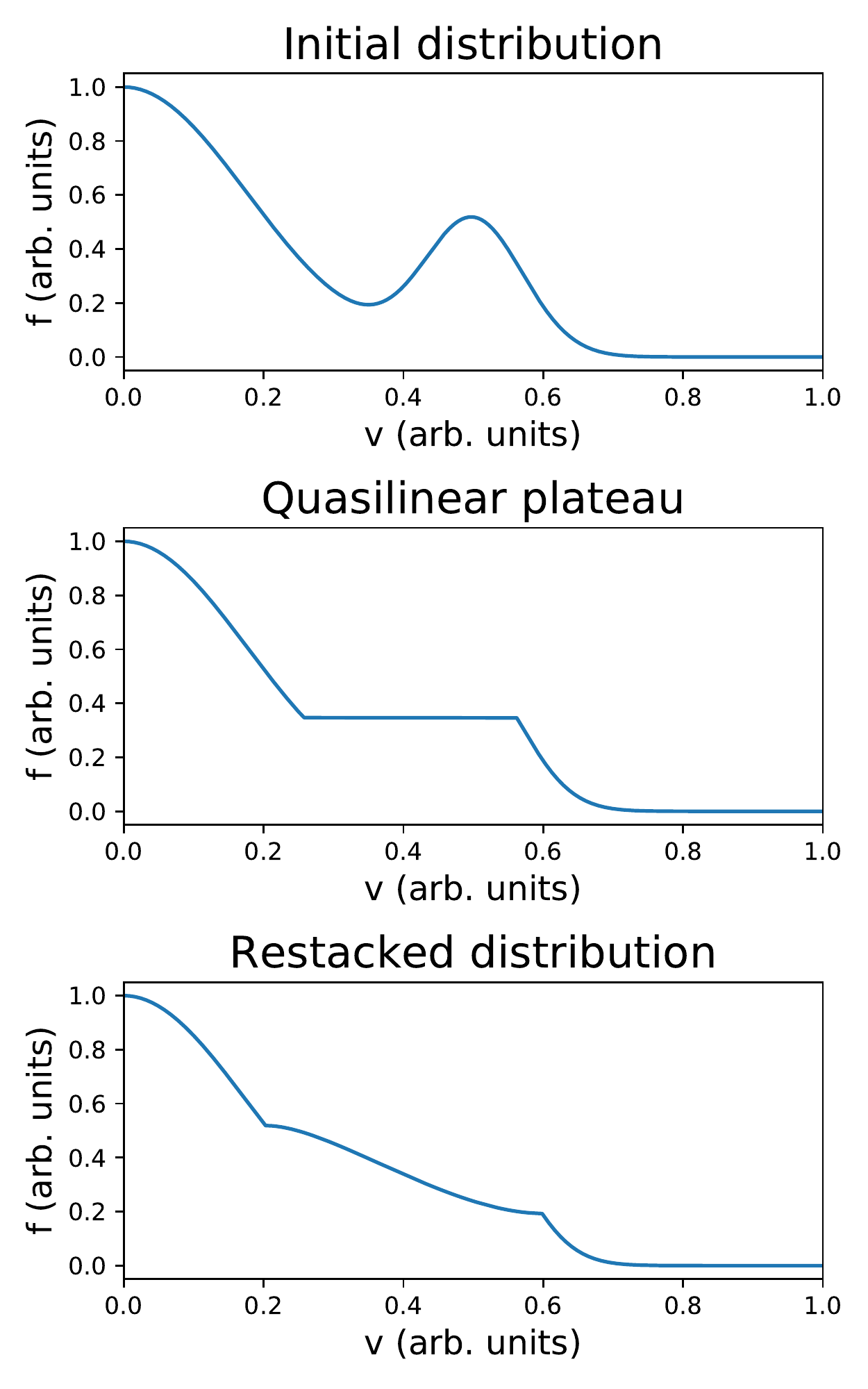}
	\caption{Top panel: classic bump-on-tail distribution; Middle panel: minimum energy state under diffusion exhibiting quasilinear plateau; Bottom panel: minimum energy state under Gardner restacking.} \label{fig:bumpOnTail}
\end{figure}

\section{Bump-on-Tail Distribution} \label{sec:bumpOnTail}

Consider the classic bump-on-tail instability, which features an initial distribution with nonzero free energy (top panel: Figure~\ref{fig:bumpOnTail}).
Bump-on-tail distributions are unstable to interactions with waves. 
This instability, and its saturation, are major results in the theory of quasilinear diffusion \cite{KrallTrivelpiece}. 

Consider first the nature of the minimum energy state under Gardner restacking. 
Interestingly, Gardner restacking does not, in general, map smooth distributions to smooth distributions. 
For smooth initial distributions, restacking  preserves uniform continuity \cite{Baernstein}, but it generates discontinuities in the derivatives that correspond to local extrema of the initial distribution (bottom panel: Fig.~\ref{fig:bumpOnTail}). 
This makes intuitive sense. 
For instance, if one divides an initial one-dimensional distribution $f(x)$ into monotonic segments, a given segment will have no influence on the restacked distribution above its maximum or below its minimum but can abruptly become important at these values. 
This can generate discontinuities in the derivatives of the restacked $f$, but not discontinuities in the restacked $f$ itself. 

Compare now the minimum energy state under Gardner restacking with the minimum energy state under diffusion. 
The classic minimum energy state  for the bump-on-tail distribution (middle panel: Fig.~\ref{fig:bumpOnTail}) assumes only local diffusion, leading to a constant density region that allows the velocity-space  {\it bump} to  fill in the contiguous,  lower-energy, velocity-space {\it valley}. 
The restacking operations clearly release considerably more energy than do the local diffusive operations.

However, in principle,  should it be possible to arrange waves so as to perfectly control the diffusion paths in phase space, then the diffusion need not be local.  
For example, one could imagine an arbitrarily thin  diffusion path in phase space that connects two disjoint regions,  so that particles can be diffused  between two disjoint regions without affecting the phase space between them. 
In that case, in the continuous limit,  the bump-on-tail distribution can instead be transformed to the restacked distribution (bottom panel: Fig.~\ref{fig:bumpOnTail}). 
Of course, in practice, actually transforming a bump-on-tail distribution into the restacked distribution  using waves would require an extraordinary degree of control over the waves in the system. 

In principle, the minimum energy state allowing arbitrary diffusion can be found by discretizing the phase space and searching all possible diffusive operations.  
However, finding the optimal diffusion paths, and the sequence in which they would be used, would be a prohibitively expensive search (NP hard),  
with a search space for $N$ discrete elements that has an $\mathcal{O}(N^{N^2})$ upper bound \cite{Hay2015}. 
On the other hand, the discrete restacking problem is a sorting procedure that can be completed in $\mathcal{O}(N \log N)$ operations, and thus can  approximate the continuous problem in a tractable way.

\section{Discussion} \label{sec:discussion}

The key result  here is the demonstration that, in the continuous limit, the free energy accessible by diffusive exchanges between phase space volumes is exactly the Gardner free energy. 
Moreover, a prescription of diffusive operations is constructed for unlocking this energy.
In so doing, we reach the curious result that macroscopic phase-space conservation can be achieved by ostensibly entropy-producing microscopic operations. 

Whether or not the constructed sequence is the most efficient sequence of steps, the fact that it  leads to a release of energy that converges to the Gardner restacking limit provides an upper bound to the number of steps to extract the Gardner free energy. 
Specifically, it is shown (in Section~\ref{sec:sequenceProof})  that when each discrete region is divided into $N$ smaller regions, the fractional difference between the diffusive free energy and the Gardner free energy vanishes at least as quickly as $\mathcal{O}(N^{-1/2})$. 
It remains unresolved whether the optimal sequence of diffusive exchanges might scale more quickly.

The ability of diffusive exchanges operating on a finer scale to extract the Gardner free energy on a coarser scale may be thought of as the converse to the recognition that ostensibly entropy-preserving, fine-grain, reversible operations can appear to be diffusive when viewed on a coarse scale.  
Instead, as shown here, a reversible exchange between two phase space volumes can be constructed out of many finer-grained diffusive exchanges. 

This converse now provides insight into how operational constraints can be circumvented by finer granulation of the phase space to release energy approaching the Gardner restacking limit. 
Similar behavior can result from increasing the volume of the accessible phase space, as can happen when additional conservation laws are relaxed \cite{Kolmes2020ConstrainedDiffusion}. 

The fact that macroscopic phase-space conservation can be achieved by entropy-producing microscopic operations has consequences beyond the release of free energy.  
Any objective function, not necessarily the free energy, can be optimized in a similar way through diffusive operations if those operations take place on a fine enough scale.  
In other words, whatever can be accomplished by restacking can also be accomplished, in principle, by diffusion.
For example, just as Gardner restacking optimized energy extraction by assigning the highest density phase space volumes to the lowest energies, so too could the fusion reactivity be maximized, in principle, by assigning the highest density phase space volumes to the energies closest to where the fusion cross-section is maximal.
A second, perhaps more practical example is the optimal rearrangement of the six dimensional electron phase space so as to support the most electric current while incurring the least power dissipation. This would be a generalization of driving currents by diffusion of particles by specific RF waves \cite{Fisch1987}, but with the diffusive operations covering the full distribution function. 

Finally, it is noteworthy that there are deep analogies between recent advances in structure-preserving algorithms \cite{Qin2013, He2015, Morrison2017, Glasser2020} and in the use here of entropy-producing microscopic operations to produce entropy-preserving macroscopic behavior. Some structure-preserving algorithms, which feature conservation properties that are important for the reliability of long-time solutions, employ splitting algorithms, which have similarities to the procedure described here. It can be hoped, therefore, that the solution offered here of the continuous-limit, diffusive-exchange problem might also provide useful insights into these important new computational methods, which have recently been applied across a variety of areas of physics.

\begin{acknowledgments}
The authors would like to acknowledge Per Helander, Mike Mlodik, and Ian Ochs for helpful conversations. 
This work was supported by U.S. DOE DE-SC0016072 and DOE NNSA DE-NA0003871. 
\end{acknowledgments}

\providecommand{\noopsort}[1]{}\providecommand{\singleletter}[1]{#1}%

\end{document}